\documentclass{aa}
\usepackage{graphicx}
\usepackage[varg]{txfonts}
\usepackage{txfonts}
\usepackage{amsmath}

\usepackage{amsmath, bm}
\usepackage[colorlinks=true,citecolor=blue,linkcolor=blue]{hyperref}

\begin{document}

\title{Exploring the intermittency of magnetohydrodynamic turbulence by synchrotron polarization radiation}
 \author{Ru-Yue Wang
          \inst{1},
          Jian-Fu Zhang
          \inst{1,2,3}, 
          Fang Lu
         \inst{1}, 
          Fu-Yuan Xiang
          \inst{1,2} 
          }

\institute{Department of Physics, Xiangtan University, Xiangtan, Hunan 411105, People's Republic of China,
\email{jfzhang@xtu.edu.cn; flu@xtu.edu.cn}
         \and
         Key Laboratory of Stars and Interstellar Medium, Xiangtan University, Xiangtan 411105, People’s Republic of China
         \and
         Department of Astronomy and Space Science, Chungnam National University, Daejeon, Republic of Korea
             }


\abstract
{Magnetohydrodynamic (MHD) turbulence plays a critical role in many key astrophysical processes such as star formation, acceleration of cosmic rays, and heat conduction. However, its properties are still poorly understood.}
{We explore how to extract the intermittency of compressible MHD turbulence from the synthetic and real observations. }
{The three statistical methods, namely the probability distribution function, kurtosis, and scaling exponent of the multi-order structure function, are used to reveal
the intermittency of MHD turbulence.}
{Our numerical results demonstrate that: (1) the synchrotron polarization intensity statistics can be used to probe the intermittency of magnetic turbulence, by which we can distinguish different turbulence regimes; (2) the intermittency of MHD turbulence is dominated by the slow mode in the sub-Alfv{\'e}nic turbulence regime; (3) the Galactic  interstellar
medium (ISM) at the low latitude region corresponds to the sub-Alfv\'enic and supersonic turbulence regime.}
{We have successfully measured the intermittency of the Galactic ISM from the synthetic and realistic observations.}


\authorrunning{Wang et al.}
\titlerunning{Exploring the intermittency by synchrotron polarization}
\maketitle

\section{Introduction}
Intermittency is one of the properties of MHD turbulence, which has been studied in the solar wind \citep{Veltri1999, Carbone2004, Chen2014, Wang2015}, interplanetary medium \citep{Bruno2007, Zelenyi2015}, magnetosphere \citep{Xu2023} and interstellar medium \citep{McKee1977, Rickett2011, Falgarone2011, Fraternale2019}. The intermittency, associated with a small-scale coherent structure, can influence interstellar gas heating \citep{Osman2011, Osman2012a,Osman2012, Wu2013,Chen2020, Phillips2023}, energy dissipation \citep{Zhdankin2016, Wan2012, Huang2022}, and increased temperature anisotropy \citep{Servidio2012, Osman2012} in plasma turbulence. Moreover, this coherent structure plays an important role in particle acceleration \citep{Decamp2006, Lemoine2021, Vega2023} and scattering \citep{Butsky2023}. Therefore, studying intermittency is significant for understanding and interpreting several astrophysical processes. 
 
When neglecting the intermittency effect, \citet{K41} predicted that the power-law index $\zeta_{p}$ of structure function of velocity fluctuations is proportional to its order $p$ in the inertial range, namely $\zeta_{p}=p/3$. The intermittency is ubiquitous in turbulence environments such as the solar wind (e.g., \citealt{Osman2014}) and diffusion ISM (e.g., \citealt{Falgarone2011}). With intermittency, the relation between the order and the power-law index is expected to be a non-linear behaviour. Indeed, the later studies found that the scaling exponent $\zeta_{p}$ gradually deviates from linear relation of $\zeta_{p}=p/3$ as the order $p$ increases \citep{Anselmet1984, Meneveau1987, bialas1990, Vincent1991}. To understand this phenomenon, two modified models have been proposed to describe incompressible hydrodynamic \citep{She1994} and MHD turbulence \citep{Muller2000}, respectively (see Section \ref{sec:intermittency_theory_model}, and also \citealt{biskamp2003, Beresnyak2019} for more details).

The intermittency of MHD turbulence has been extensively investigated in numerical simulations \citep{Falgarone2008, biskamp2003, Esquivel2010}. 
For the incompressible homogeneous MHD turbulence, \citet{Yoshimatsu2011} concluded that the magnetic field is more intermittent than the velocity, consistent with other works \citep{Cho2003ApJ, Haugen2004, Mallet2017}. For the compressible MHD turbulence, \citet{Kowal2007} 
claimed that the density intermittency strongly depends on Alfv\'enic and sonic Mach numbers, with the velocity intermittency being different from density one. 

The later simulations confirmed that the intermittency of MHD turbulence is closely related to its compressibility, scale-dependent anisotropy, and magnetization. For instance, it was found that the intermittency of three plasma modes (Alfv\'en, slow, and fast) increases with the sonic Mach number in the weak magnetic field (\citealt{Kowal2010}). The viscosity-damped MHD turbulence shows the scale-dependent intermittency, i.e., the more intermittency at much smaller scales (\citealt{Cho2003ApJ}). For the latter, \citet{Davis2023} concluded that with increasing magnetization, the velocity fluctuations display an inverse trend of the co-dimension of structures compared to the magnetic field.
\citet{Yang2018} also presented that the effect of MHD turbulence amplitude on the distribution of magnetic field is preferred to that of the angle to the local magnetic field. Besides, it is claimed that the intermittency is associated with the driving way of turbulence \citep{Federrath2008ApJ, Federrath2009ApJ, Beattie2022}. Based on statistics of centroid velocity, \citet{Federrath2010} found that the intermittency for compressive forcing is stronger than that for solenoidal one.

Note that there are studies on measuring MHD turbulence intermittency in solar physics and astrophysics. For the former, with the solar wind data from the STEREO spacecraft, \citet{Osman2014} measured the intermittency of magnetic and Els{$\ddot{a}$}sser field fluctuations in the solar wind and found that the intermittency in the direction perpendicular to the local magnetic field is stronger than that in the parallel direction. Considering density fluctuations, \citet{Chen2014} also found strong intermittency ranging from ion to electron scales. For the latter, the turbulence in molecular clouds exhibits small-scale and inertial-range intermittency \citep{Hily-Blant2008}. In addition, \citet{Falgarone2011} found the non-Gaussian statistic results and the existence of coherent structures in the diffuse interstellar medium (ISM).

As mentioned above, the measurement of intermittency has been performed using spectroscopic data of ISM and in situ data of the solar wind. Can we use synchrotron polarization observations to extract the intermittency of MHD turbulence? One purpose of our numerical studies is to explore the intermittency of magnetic fields and densities via synthetic observations. Another purpose is to study the intermittency of the Galactic ISM using realistic observations. Specifically, we first synthesize synchrotron polarization observations using numerical simulation data to study the intermittency of the magnetic field and density, and then adopt realistic observations from the Canadian Galactic Plane Survey (CGPS) to explore the intermittency of the Galactic ISM. From an observational perspective, this work will be dedicated to advancing the understanding of the intermittency of compressible MHD turbulence.

The structure of this paper is organised as follows. In Section \ref{sec:theory}, we give descriptions of theoretical models of intermittency, synchrotron radiative processes, and methods to characterise intermittency. Section \ref{sec:numerical_simulation} introduces numerical setup and decomposed method of compressible MHD turbulence. Section \ref{sec:numerical_results} presents numerical results, followed by the studies of intermittency from the observational data in Section \ref{sec:observation_results}.
The discussion and summary are provided in Sections \ref{sec:discussion} and \ref{sec:conclusion}, respectively. 

\section{Theories and methods} \label{sec:theory}
\subsection{Theoretical models related to intermittency} \label{sec:intermittency_theory_model}

In the framework of incompressible hydrodynamic turbulence, \cite{K41} assumed that the turbulence is self-similar within the inertial range, in which the relationship between the velocity fluctuations ${\delta u}_{\rm l}$ and the scale $l$ exhibits a simple scaling behaviour as follows
\begin{equation}
\langle {{\delta u}_{\rm l}^{p}} \rangle \sim l^{\zeta_{p}}, ~ {\zeta_{p}}=\frac{p}{3}.
\end{equation}
When the scaling exponent $\zeta_{\rm p}$ deviates from this relation, it will imply the appearance of intermittency phenomenon, which reflects the inhomogeneous distribution of fluctuations \citep[see Chap.~7 in] []{biskamp2003}. Later, taking the scaling of velocity $u_{\rm l} \sim l^{1/g}$ and energy cascade rate $t^{-1}\sim l^{-x}$ into account, \cite{She1994} analytically proposed a classical non-linear scaling and expressed by  
\begin{equation}
\zeta_{p}=\frac{p}{g}(1-x)+C(1-(1-x/C)^{p/g}), \label{scaling_exponent}
\end{equation}
where $C$ denotes the co-dimension of the dissipative structures related to the dimension of a dissipative structure $D$ via the relation of $C=3-D$.

In the case of hydrodynamic turbulence, one usually considers the parameters $g=3$ and $x=2/3$ (according to Kolmogorov scaling). For the 1D vortex filament ($C=3-D=2$), Equation (\ref{scaling_exponent}) can be simplified as
\begin{equation}
\zeta_{p}=\frac{p}{9}+2[1-(2/3)^{p/3}], \label{SL} 
\end{equation}
which is called the She \& Leveque (SL) model in this paper.
For the 2D sheet-like structure, it can be rewritten as \citep{Muller2000}
\begin{equation}
\zeta_{p}=\frac{p}{9}+1-(1/3)^{p/3}, \label{MB}
\end{equation}
which is called the M\"{u}ller \& Biskamp (MB) model.

\subsection{Synchrotron radiative processes} \label{sec:synchrotron_radiation}

The production of synchrotron radiation requires two key factors, namely the relativistic electrons and the magnetic field. In this paper, we assume that the relativistic electron population follows isotropic pitch-angle distribution and has the following power-law relationship 
\begin{equation}
N(E)dE=N_{0}E^{2\alpha-1}dE,   \label{electron distribution}
\end{equation}
where $N(E)dE$ is the number density of relativistic electrons in the energy interval $E$ and $E+dE$, $N_{0}$ a normalisation constant, and $\alpha=(1-p)/2$ the photon spectral index related to the electron index $p$. In the simulation below, we set the photon spectral index of $\alpha=-1.0$ for simplicity.

The synchrotron radiation intensity is expressed as \citep{Ginzburg1965}
\begin{equation}
I(\bm{X}) \varpropto \int_{0}^{L} B_{\perp}^{1-\alpha}({\bm X}, z) dz,
\end{equation}
where $B_{\perp}$ is the component of the magnetic field perpendicular to the line of sight (LOS), ${\bm X}=(x, y)$ a 2D vector in the plane of the sky (POS), and $L$ the spatial length of emitting region. 

Considering the linearly polarized properties of synchrotron radiation, we have the intrinsic polarization intensity 
\begin{equation}
P_{0}({\bm X})=p_{0}I({\bm X}), 
\end{equation}
where $p_{0}=(3-3\alpha)/(5-3\alpha)$ is the fraction polarization degree.  
The observable Stokes parameters $Q$, $U$ can be expressed as $Q({\bm X})=P_{0}({\bm X})\cos2\phi$ and $U({\bm X})=P_{0}({\bm X})\sin2\phi$, respectively. Here, the angle $\phi=\phi_{0}=\pi/2+\arctan(B_{\rm y}/B_{\rm x})$ represents the polarization angle. When involving Faraday rotation effect, this angle can be expressed as $\phi=\phi_{0}+\lambda^{2}\varphi$, with Faraday rotation measure (RM) of $\varphi({\bm X}, z)=0.81\int_{0}^{z}n_{\rm e}({\bm X}, z^{'}) B_{\parallel}({\bm X}, z^{'}) dz^{'} \rm rad~ \rm m^{-2}$, where $n_{\rm e}$ represents the number density of thermal electrons and $B_{\parallel}$ the component of the magnetic field along the LOS.
Defining the complex polarization vector of ${\bm P}=Q+iU$, we have the synchrotron polarization intensity (SPI) of 
\begin{equation}
P=\sqrt{Q^2+U^2} \label{equ:synchrotron_polarization}.
\end{equation}

\subsection{Methods to characterise intermittency} \label{sec:method}
Firstly, the appearance of intermittency can be revealed by the probability distribution function (PDF). To characterise the statistical behaviour at a specific separation $\bm R$, we can calculate the PDF of the dispersion $\delta F(\bm R)$, which is defined as 
\begin{equation}
\delta F(\bm R)= F({\bm X}+{\bm R})-F({\bm X})
\end{equation}
for any fluctuation quantity $F$, where ${\bm X}$ represents a 2D position vector in the POS. In general, the PDFs of fluctuations exhibit a non-Gaussian distribution with two extended tails when the intermittency occurs.

Secondly, to further understand the intermittency over the whole spatial scale, we need to use another method such as kurtosis and scaling exponent. The multi-order structure function can be defined as 
\begin{equation}
{\rm SF}_{p}({\bm R})=\langle | F({\bm X}+{\bm R}) - F({\bm X})|^{p} \rangle,
\end{equation}
where $\langle ...\rangle$ represents a spatial average of the system. Using the second- and fourth-order structure functions, we can define the kurtosis as \citep{Bruno2003}
\begin{equation}
{K}=\frac{{\rm SF}_{4}(\bm R)}{({\rm SF}_{2}(\bm R))^{2}},
\end{equation}
the value of which will reflect the distribution of fluctuations. If ${K}\neq 3$, the fluctuations have a non-Gaussian distribution. In addition, the $K$ change with $\bm R$ can characterise the level of intermittency \citep{Frisch1995}. When $K$ grows faster, the fluctuations are more intermittent. When $K$ remains a constant within a certain scale range, the fluctuations are self-similar and not intermittent.

Thirdly, intermittency can also be measured by the scaling exponent of the multi-order structure function. The multi-order structure function is related to the separation scale ${\bm R}$ within the inertial range, and described by a power-law relation of
\begin{equation}
{\rm SF}_{p}({\bm R})\varpropto {\bm R}^{\zeta(p)},
\end{equation}
where $\zeta (p)$ is the absolute scaling exponent related to the order of structure function.
In this paper, we adopt the extended self-similarity hypothesis \citep{Benzi1993}, that is, the power-law scaling can be extended from the inertial range to the dissipation scale.
Under this hypothesis, we explore the scaling exponent $\xi(p)$ between the $3$rd- and $p$th-order structure functions, by which we distinguish the intermittency level.
When the relation between the scaling exponent $\xi(p)$ and the order $p$ is nonlinear, it represents the presence of intermittency with the multi-fractal feature.

\section{MHD Turbulence Simulation} \label{sec:numerical_simulation}
The second-order-accurate hybrid essentially non-oscillatory code \citep[see][]{Cho2003mnras} is used to solve the ideal single-fluid MHD equations (i.e., only including the proton component $\rho$ to simulate MHD turbulence)
\begin{align}
\frac{\partial \rho }{\partial t} + \nabla \cdot (\rho {\bm v})=0, \label{eq:den}\\
\rho[\frac{\partial {\bm v}}{\partial t} + ({\bm v}\cdot \nabla) {\bm v}] +  \nabla p_{\rm g}- \frac{{\bm J} \times {\bm B}}{4\pi} ={\bm f}, \label{eq:vel}\\
\frac{{\partial {\bm B}}}{{\partial t}} -\nabla \times ({\bm v} \times{\bm B})=0,\label{eq:mag}\\
\nabla \cdot {\bm B}=0,
\end{align}
where $t$ is the evolution time of turbulence, $p_{\rm g}=c_{\rm s}^{2}\rho$ the thermal gas pressure, ${\bm J}=\nabla \times {\bm B}$ the current density, and ${\bm f}$ a random driving force. These physical quantities are dimensionless. The computation domain is a cube with a side length of $2\pi$. Periodic boundary conditions are applied at the computational boundaries.

Using a numerical resolution of $512^3$, we drive the turbulence by a solenoidal driving force acting on the wavenumber of $k\approx2.5$, with a continuous injection of energy. We use a three-stage Runge-Kutta method for time integration, in units of the large eddy turnover time of $\sim L/\delta V$. Meanwhile, we also set the initial magnetic field ($B_{\rm init}$) along the $x$ axis and the gas pressure ($P_{\rm init}$). To characterise different models, we define three parameters: Alfv{\'e}nic Mach number $M_{\rm A}={V_{\rm L}}/{V_{\rm A}}$, sonic Mach number $M_{\rm s}={V_{\rm L}}/{c_{\rm s}}$ and plasma parameter $\beta=2{M_{\rm A}}^{2}/{M_{\rm s}}^{2}$, where $V_{\rm L}$ is the injection velocity, and $V_{\rm A}=B_{\rm init}/\sqrt{4\pi\rho}$ is the Alfv{\'e}nic velocity. The first two parameters characterise the strength of magnetic field and compressibility, respectively. The latter indicates the ratio of thermal to magnetic pressure, in which the magnetic field is dynamically important ($\beta<1$) or unimportant ($\beta>1$). The related parameters are listed in Table \ref{table:1}.

Based on data cubes, we decompose compressible MHD turbulence into three modes in Fourier space, the unit vectors of which are defined by (\citealt{Cho2002PhRvL})
\begin{align}
{\hat \Xi}_{\rm f} \propto (1+\frac{\beta}{2}+\sqrt{D})(k_{\perp} {\bm{\hat k}_{\perp}})+(-1+\frac{\beta}{2}+\sqrt{D})(k_{\parallel} {\bm{\hat k}_{\parallel}}), \label{eq:fast}\\
{\hat \Xi}_{\rm s} \propto (1+\frac{\beta}{2}-\sqrt{D})(k_{\perp} {\bm{\hat k}_{\perp}})+(-1+\frac{\beta}{2}-\sqrt{D})(k_{\parallel} {\bm{\hat k}_{\parallel}}), \label{eq:slow}\\
{\hat \Xi}_{\rm A} \propto -{\bm {\hat{k}}}_{\perp} \times {\bm{\hat{k}}}_{\parallel},\label{eq:alfven}
\end{align}
with $D=(1+\frac{\beta}{2})^{2}-2\beta \cos^{2}\theta$ and $\cos\theta={\bm{\hat k}}_{\parallel} \cdot {\bm {\hat{B}}}$. When projecting the magnetic field onto these unit vector directions, we obtain the magnetic field components of each mode in Fourier space. These projection quantities are then transformed into a real space to recover the corresponding magnetic field. 

\begin{table*}
\caption{Different models of compressible MHD turbulence.}
\centering
\setlength{\tabcolsep}{5mm}
\begin{tabular}{cccccccc}
\hline\hline
Models  & {$B_{\rm init}$} & {$P_{\rm init}$} & {$M_{\rm A}$} & {$M_{\rm s}$} & {$\beta$} & {$\delta B_{\rm rms}/B_{\rm init}$} & {Descriptions}\\
\hline
  Run1  & 1.0 & 2.0 & 0.65  & 0.48  & 3.668  & 0.614 & Thermal pressure dominated\\
  Run2  & 1.0 & 0.025 & 0.55  & 4.46  & 0.030  & 0.467 & Magnetic pressure dominated\\
  Run3  & 0.1 & 2.0 & 1.72  & 0.45  & 29.219  & 6.345 & Thermal pressure dominated\\
  Run4  & 0.1 & 0.05 & 1.69  & 3.11  & 0.591  & 5.254 & Magnetic pressure dominated\\ 
\hline
\end{tabular}
\tablefoot{$B_{\rm init}$--- magnetic field strength; $P_{\rm init}$--- gas pressure; $M_{\rm A}$ --- Alfv{\'e}nic Mach number; $M_{\rm s}$---sonic Mach number; $\beta$---plasma parameter; $\delta B_{\rm rms}$---root mean square of the random magnetic field. $B_{\rm init}$ and $P_{\rm init}$ are our initial parameter setting, respectively. The other resulting values are obtained from the final snapshot data.} \label{table:1}
\end{table*}

\begin{figure*}
\centering
\includegraphics[width=1.0\textwidth]{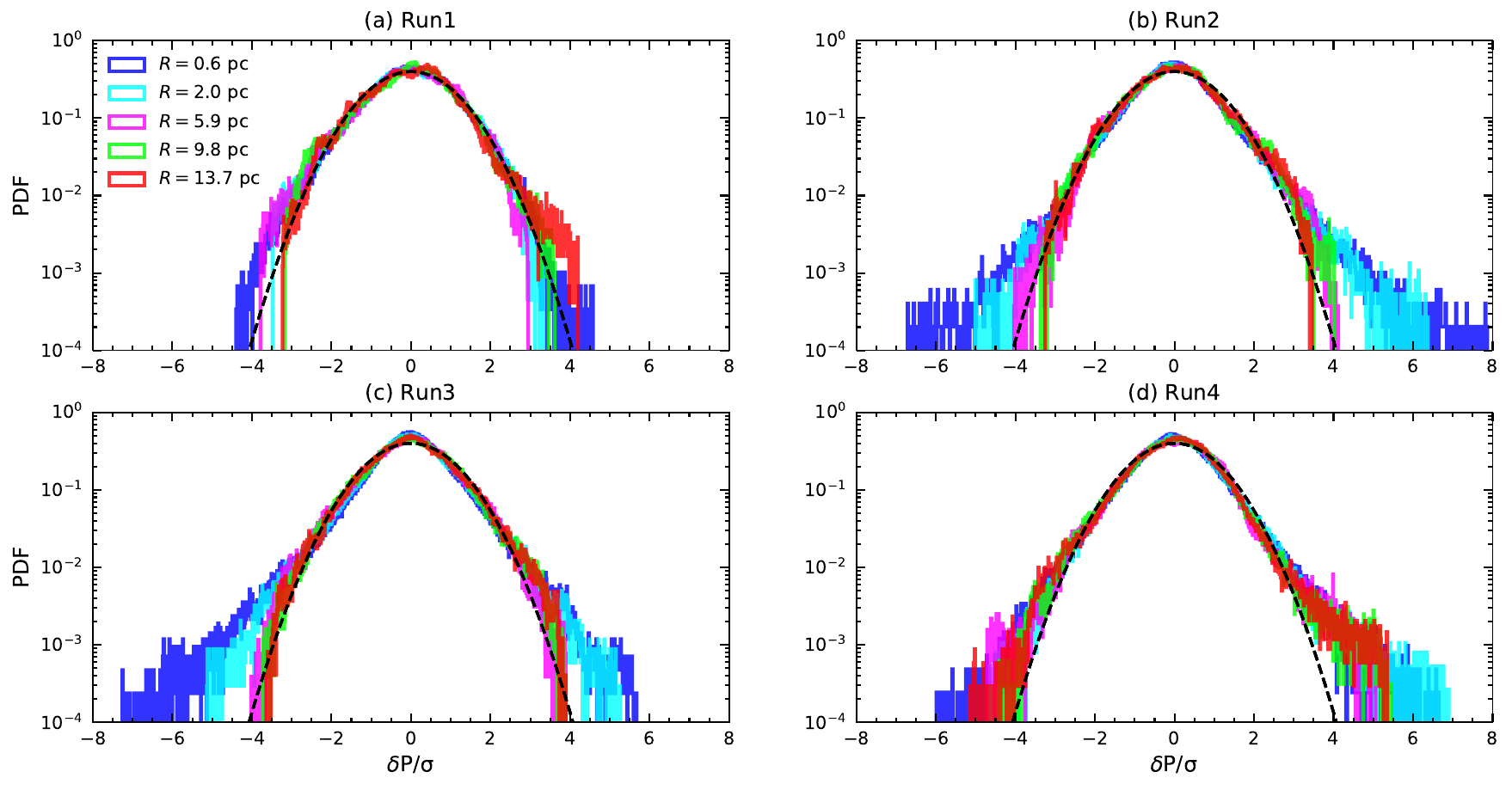}
\caption{The PDFs of SPI normalised by its standard deviation $\sigma$ at different scales $R$, arising from four different turbulence regimes. The black dashed lines represent the Gaussian distributions.
}\label{fig:synchrotron_pdf_four_regime} 
\end{figure*}

\section{Synchrotron Polarization Simulation} \label{sec:numerical_results}

\subsection{The measurement of intermittency arising from different turbulence regimes}

To generate synthetic observations, we calculate the SPI via Equation (\ref{equ:synchrotron_polarization}) using the above data cubes, with the assumption of the thermal electron density $n_{\rm e}$ proportional to the plasma density $\rho$, i.e., $n_{\rm e}=\rho$ when involving Faraday rotation measure.
We use the typical values of the Galactic ISM to parameterise dimensionless physical quantities. Here, we just provide three key parameters such as the spatial length of $L=100~\rm pc$ along the LOS, the thermal electron density $n_{\rm e}=0.1~\rm cm^{-3}$, and the magnetic field strength $B=1.23~\rm \mu G$. With the numerical resolution of $512$ pixels, we have a mesh grid of $100~{\rm pc}/512\sim 0.2~{\rm pc}$, corresponding to the smallest resolved spatial length.


We first analyse the PDFs of SPI arising from four turbulence regimes (see Table \ref{table:1}). The resulting finding is shown in Fig. \ref{fig:synchrotron_pdf_four_regime}, from which we see that PDFs at different scales $R$ exhibit other characteristics that deviate from the Gaussian distribution. In general, this deviation mainly occurs in the two tail parts of the Gaussian distribution. With the decreasing scale $R$, we find that the level of deviation from both tails increases (except for panel (a)), indicating intermittent enhancement. Comparing all four scenarios, the PDFs can qualitatively reveal the presence or disappearance of intermittency. Next, we will quantitatively evaluate the intermittency level using the kurtosis.

\begin{figure}
\includegraphics[width=0.45\textwidth]{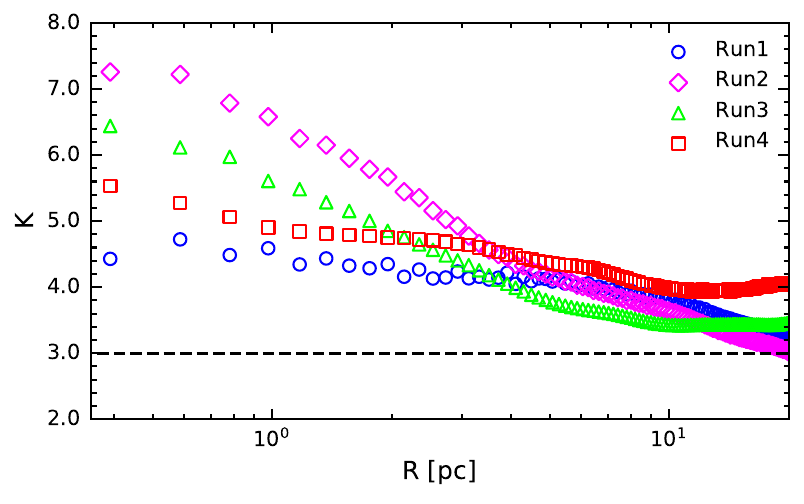}
\caption{The kurtosis of SPI as a function of the separation scale $R$ in different turbulence regimes. The horizontal dashed line corresponds to the kurtosis values of the Gaussian distribution. }\label{fig:synchrotron_k_four_regime}
\end{figure}

The kurtosis distributions of SPI over the separation scale $R$ are presented in Fig. \ref{fig:synchrotron_k_four_regime}, where the horizontal dashed line corresponds to $K=3$ representing the kurtosis value of Gaussian distribution shown in Fig. \ref{fig:synchrotron_pdf_four_regime}. As shown in Fig. \ref{fig:synchrotron_k_four_regime}, the kurtosis values of Run2 and Run3 decrease faster than those of Run1 and Run4 as the separation scale increases. This reflects the more obvious intermittency of both Run2 and Run3 at small scales. In other words, the greater the deviation from $K=3$, the more the intermittency. The kurtosis of SPI shows an irregular coupling between the Mach numbers $M_{\rm A}$ and $M_{\rm s}$. Although there is intermittency at a small scale, we cannot find a significant correlation between the kurtosis distribution and Mach numbers. However, it is apparent that the most intermittency corresponds to the largest deviations of $\beta$ from unity.

\begin{figure*}
\centering
\includegraphics[width=1.0\textwidth]{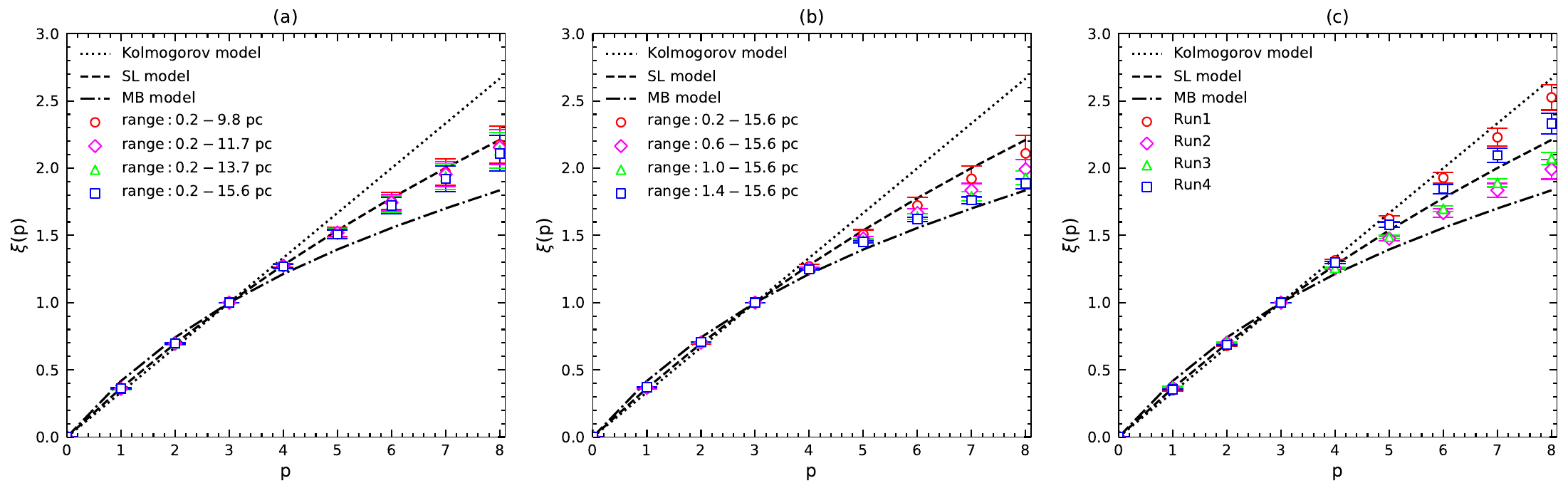}
\caption{The scaling exponent as a function of the order for the SPI at three scenarios: the lower limits of the fixed $R$ (panel (a)), the upper limits of the fixed $R$ (panel (b)), and different turbulence models (panel (c)). The results of panels (a) and (b) are obtained by Run2.
}\label{fig:synchrotron_scaling} 
\end{figure*}

In addition to the methods discussed above, the scaling exponent of multi-order structure functions is the third method for measuring intermittency. Specifically, we use the extended self-similarity \citep{Benzi1993} to obtain the scaling exponent between the 3rd- and $p$th-order structure function. Here, we first explore how different fitting ranges for $R$ affect the scaling exponent. Our results are shown in Fig. \ref{fig:synchrotron_scaling} (a) describing the relation between the scaling exponent and the order at different upper limits of the inertial range. In this figure, the dotted, dashed, and dash-dotted lines indicate theoretical results provided by the Kolmogorov, SL, and MB models, respectively. The error bar represents the standard deviation.
From panel (a), we know that different upper limits of the expected inertial range have little effect on the measurement of the scaling index, thus we fix the separation scale $R=15.6$ $\rm pc$ as an upper limit in Fig. \ref{fig:synchrotron_scaling} (b). Similarly, Fig. \ref{fig:synchrotron_scaling} (b) explores the influence of different lower limits of the approximate dissipation scale on the scaling index. It is clear that the distribution of $\xi(p)$ with $p$ behaves similarly except for the results in the range of $0.2-15.6~\rm pc$. This may be affected by the numerical dissipation.

Based on the above exploration, we fix the lower and upper limits of spatial scales as $R=0.6$ and $R=15.6$ $\rm pc$, respectively. As is shown in Fig. \ref{fig:synchrotron_scaling} (c), the scaling exponents in different turbulence regimes have different deviations from the Kolmogorov model. For Run1, i.e., the sub-Alfv{\'e}nic and subsonic turbulence, the scaling exponent is close to linear, revealing weak intermittency.  For Run4, corresponding to the super-Alfv{\'e}nic and supersonic case, we see that there is a significant deviation from the Kolmogorov model at large order $p$, reflecting the presence of intermittency. For the other turbulence regimes (see Run2 and Run3), the distributions of the scaling exponent are almost close to the SL model, characterising more intermittency.

\subsection{The measurement of intermittency at different frequencies}

\begin{figure*}
\centering
\includegraphics[width=1.0\textwidth]{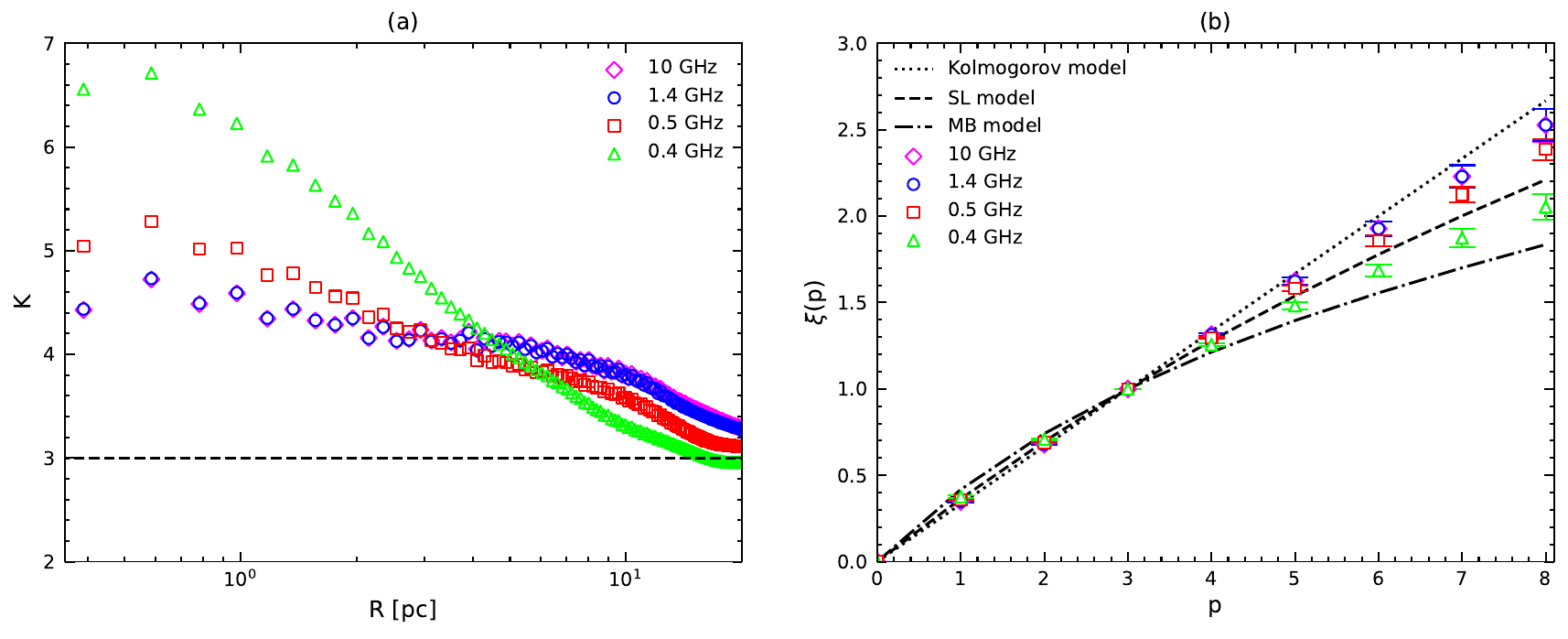}
\caption{The kurtosis (panel (a)) and scaling exponent (panel (b)) of SPI at different frequencies for the simulation of Run1. The horizontal dashed line plotted in panel (a) corresponds to the kurtosis value of Gaussian distribution. 
}\label{fig:synchrotron_k_s_four_frequency} 
\end{figure*}

Based on Run1, we explore the influence of frequency on the kurtosis and scaling exponent of SPI, respectively. The numerical results are shown in 
Fig. \ref{fig:synchrotron_k_s_four_frequency}, from panel (a) of which we see that the frequency has a significant effect on the kurtosis profiles. At low frequencies, the kurtosis shows a dramatic rise toward the small scales, while at high frequencies, the kurtosis steadily increases over the small scales. 

This reflects that the intermittency at low frequencies becomes more significant than that at high frequencies, which may be due to the Faraday depolarization effect at low frequencies making more inhomogeneous structures. It can be seen that the most significant change for kurtosis occurs at the frequency of $0.4~ \rm GHz$, indicating strong intermittency. Fig.
\ref{fig:synchrotron_k_s_four_frequency} 
 (b) shows that the scaling exponent of SPI displays different behaviours at different frequencies. At the frequency of $0.4~ \rm GHz$, there is an evident nonlinear relation close to the SL model. At the frequency of $0.5~ \rm GHz$, the scaling exponent is slight departure from the Kolmogorov model, indicating weak intermittency. Moreover, the scaling index is close to linear relation at high frequencies of 1.4 and $10~ \rm GHz$, indicating a weaker intermittency. As a result, the two methods consistently demonstrate that in the low-frequency range explored in this paper, the SPI statistics can probe the intermittency of magnetic turbulence.

\subsection{The measurement of intermittency using Faraday rotation measure}

\begin{figure*}
\centering
\includegraphics[width=1.0\textwidth]{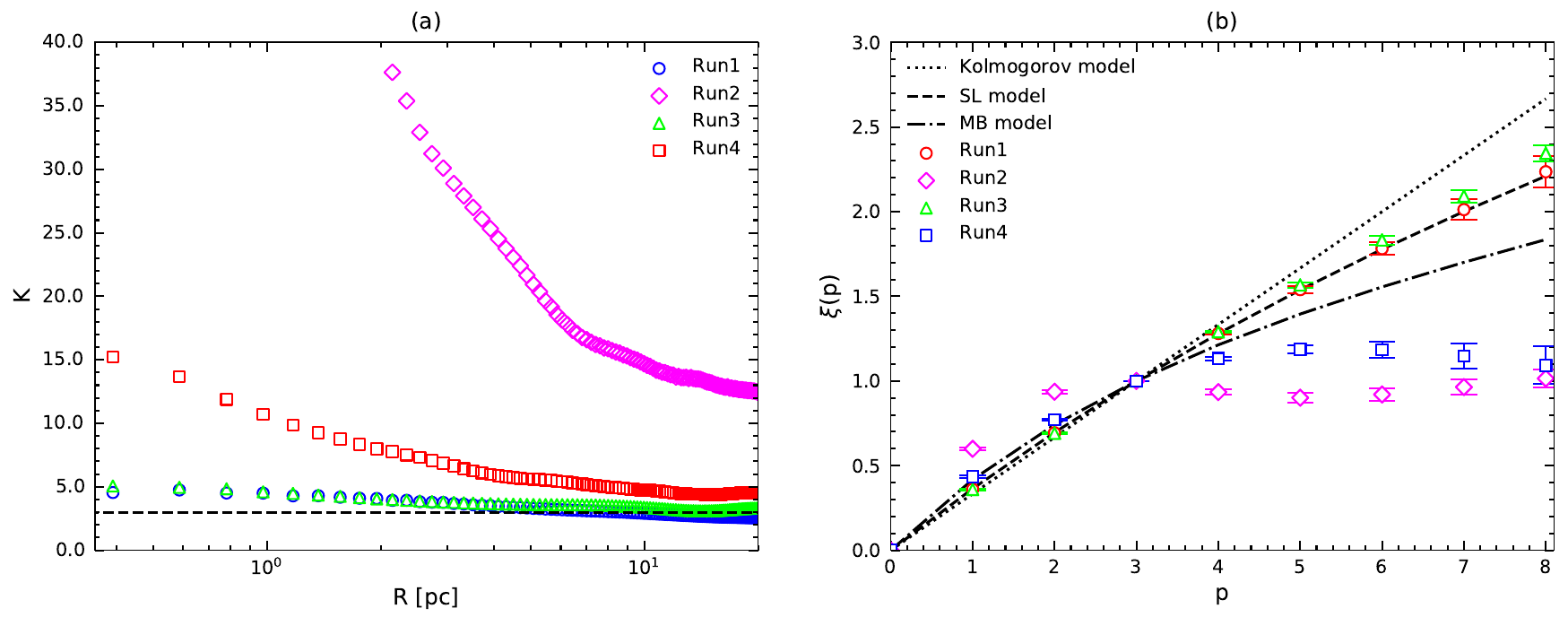}
\caption{The kurtosis (panel (a)) and scaling exponent (panel (b)) of Faraday rotation measure in different turbulence regimes. The horizontal dashed line plotted in panel (a) corresponds to the kurtosis value of Gaussian distribution.
}\label{fig:faraday_k_s_four_regime} 
\end{figure*}

The kurtosis of RMs as a function of the separation scale $R$ is presented in Fig. \ref{fig:faraday_k_s_four_regime} (a), from which we see that there are large kurtosis values of RMs at small $R$, while small values at large $R$. In addition, we also find that both supersonic turbulence cases (see Run2 and Run4) show much larger kurtosis values than subsonic ones (Run1 and Run3). This reveals that the RMs in the supersonic turbulence regime are more intermittent than those in the subsonic one. The reason may be that the formation of shocks in the supersonic turbulence increases the intermittency of MHD turbulence. Amongst four cases, kurtosis values for Run2 are the largest, exhibiting the strongest intermittency. 

Fig. \ref{fig:faraday_k_s_four_regime} (b) shows the scaling exponent for RMs as a function of the order in four different turbulence regimes. 
As shown in this panel, although all the profiles show multifractal features, the scaling exponent $\xi (p)$ varying with the order $p$ behaves differently in different sonic turbulence regimes. In the subsonic turbulence regimes, the profile of $\xi (p)$ almost follows the SL model, while it deviates far from the three theoretical models in the supersonic turbulence regimes. This reveals that the RMs in supersonic turbulence regimes are more intermittent than those in subsonic turbulence ones.  
Compared with three theoretical models, the scaling exponent in the sub-Alfv{\'e}nic and supersonic turbulence regime displays the largest deviation, which reveals the largest intermittency in this turbulence regime. Consequently, we find that the statistics of the Faraday rotation measure can recover the intermittency of MHD turbulence.

\begin{figure*}
\centering
\includegraphics[width=1.0\textwidth]{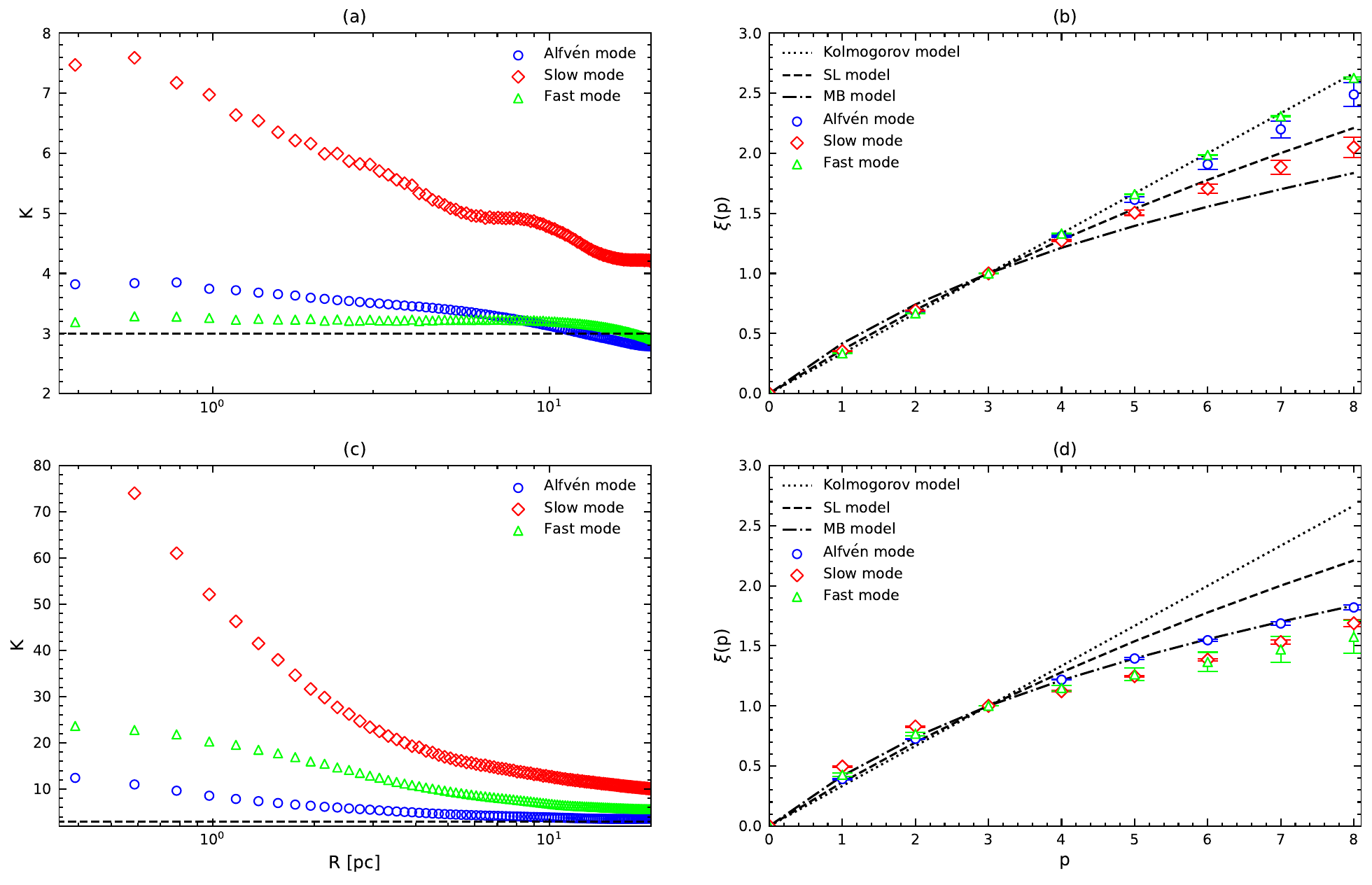}
\caption{The kurtosis (left column) and scaling exponent (right column) of SPI for three modes. The results obtained by Run1 and Run2 are shown in the upper and lower rows, respectively. The horizontal dashed lines plotted in the left column correspond to the kurtosis values of Gaussian distribution.
}\label{fig:synchrotron_three_mode} 
\end{figure*}

\subsection{The measurement of intermittency of plasma modes}

We first explore the kurtosis and scaling exponent for three plasma modes in the case of sub-Alfv{\'e}nic and subsonic turbulence (i.e., Run1). The results are shown in the upper row of Fig. \ref{fig:synchrotron_three_mode}, from panel (a) of which we see that the kurtosis of Alfv{\'e}n and slow modes decreases with the separation scale $R$, while that of fast mode remains almost unchanged. Note that the kurtosis variations of slow mode are the largest at small scales. This reflects the fact that the intermittency of SPI in the sub-Alfv{\'e}nic and subsonic turbulence regime is dominated by slow mode. From Fig. \ref{fig:synchrotron_three_mode} (b), we see that the scaling exponents for fast and slow modes follow the Kolmogorov and SL models, respectively, while for Alfv{\'e}n mode the distribution of scaling exponents lies in these two models. This may be related to the anisotropic level of three modes in the sub-Alfv{\'e}nic and subsonic turbulence regime. As demonstrated by \cite{Wang2020}, slow mode results in inhomogeneous structures because of its strong anisotropy, while fast mode produces uniform fluctuations due to its isotropy.
This suggests that slow mode has the strongest intermittency, while fast mode has no intermittency.

Moreover, the lower row of Fig. \ref{fig:synchrotron_three_mode} explores the case of sub-Alfv{\'e}nic and supersonic turbulence (i.e., Run2). Fig. \ref{fig:synchrotron_three_mode} (c) shows that the kurtosis of SPI for three modes exhibits different increasing levels as the separation scale decreases. It can be seen that the kurtosis of slow and fast modes shows a dramatic rise at small separation scales, while that of Alfv{\'e}n mode rises slowly.  
This indicates that the former two have stronger intermittency, while the latter does not manifest significant intermittency. This should be caused by the compressible nature of slow and fast modes in this turbulence regime. From Fig. \ref{fig:synchrotron_three_mode} (d), we see that the SPI for three modes displays nonlinear scaling exponents. The scaling exponents for Alfv{\'e}n mode are consistent with the MB model, while those for the other two modes deviate from this model. As a consequence of these two methods, the SPI for slow mode dominates the intermittency of MHD turbulence. In addition, compared with the results of Fig. \ref{fig:synchrotron_scaling} (c), we find that the intermittency of SPI for postdecomposition MHD modes is stronger than that for predecomposition MHD modes in the sub-Alfv{\'e}nic and supersonic turbulence regime, which may be weakened by the coupling of three modes.

\begin{figure*}
\centering
\includegraphics[width=1.0\textwidth]{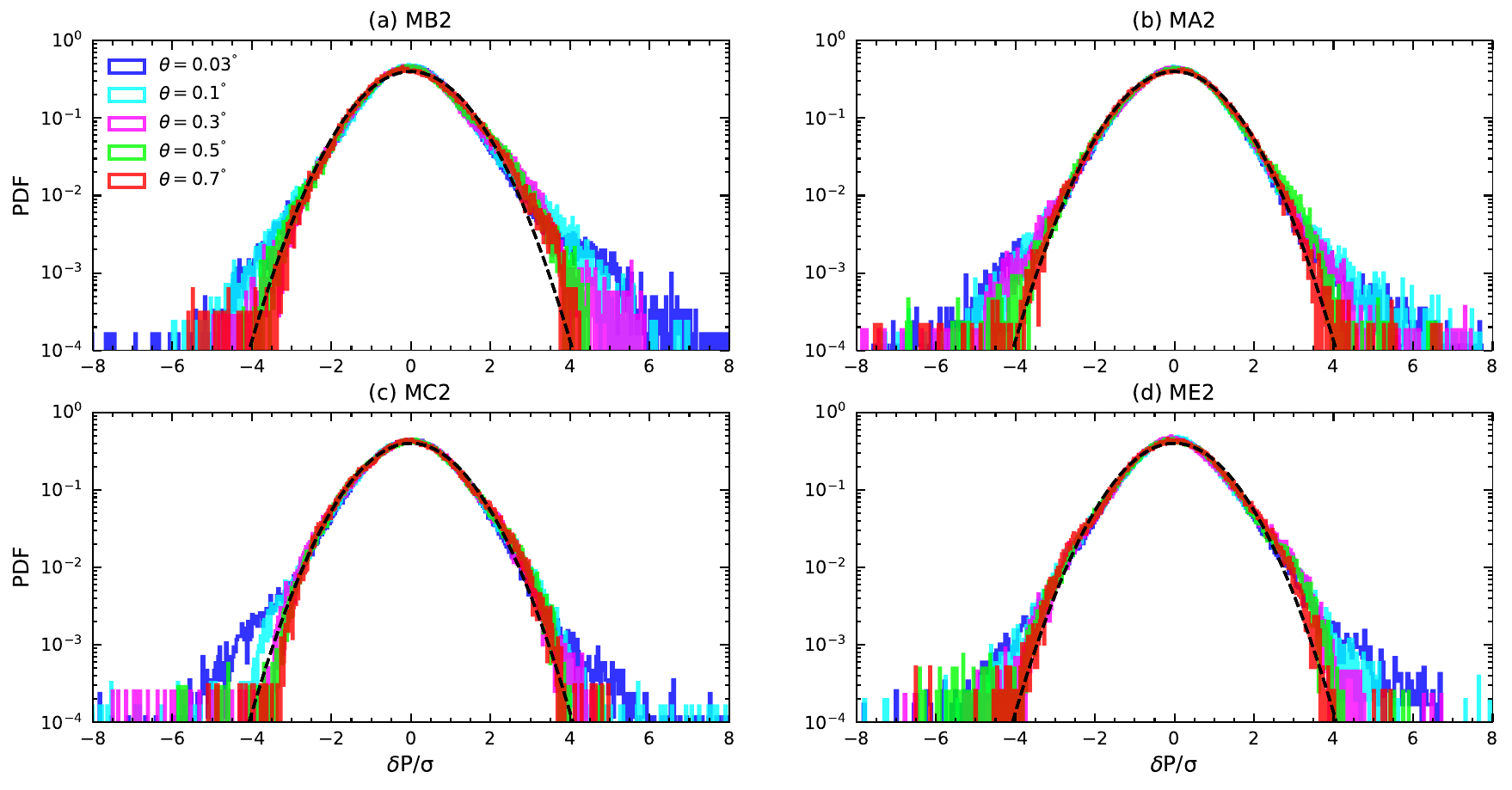}
\caption{The PDFs of SPI are normalised by its standard deviation $\sigma$ at different angles $\theta$, arising from four sets of CGPS data. The black dashed lines represent the Gaussian distributions.
}\label{fig:real_observation_PDF} 
\end{figure*}

\begin{figure*}
\centering
\includegraphics[width=1.0\textwidth]{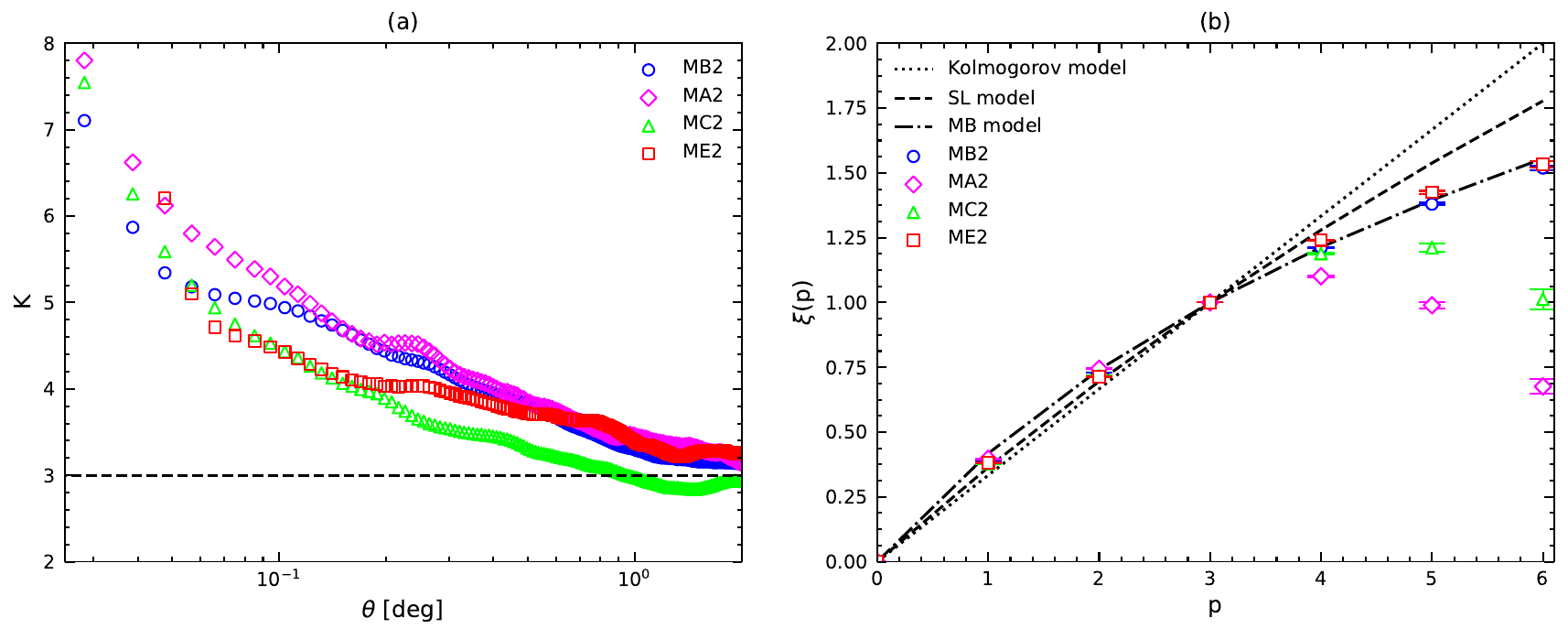}
\caption{The kurtosis (panel (a)) and scaling exponent (panel (b)) of SPI for four sets of CGPS data at 1.42 GHz. The horizontal dashed line plotted in panel (a) corresponds to the kurtosis values of Gaussian distribution.
}\label{fig:synchrotron_k_s_four_regime_CGPS} 
\end{figure*}

\begin{table}
\caption{The CGPS archive data observed at $1.42~\rm GHz$.}
\label{table:2}
\centering
\setlength{\tabcolsep}{1.8mm}
\begin{tabular}{ccc}
\hline\hline
Mosaic  & Galactic Longitude (deg) & Galactic Latitude (deg) \\
\hline
  MA2  & $131.3<l<128.8$  & $0.4<b<3.0$  \\
  MB2 & $127.3<l<124.8$  & $0.4<b<3.0$  \\
  MC2  & $123.3<l<120.8$  & $0.4<b<3.0$  \\
  ME2  & $115.3<l<112.8$  & $0.4<b<3.0$  \\  
\hline
\end{tabular}
\end{table}

\section{Application to observations} \label{sec:observation_results}
In this section, we explore the intermittency of the Galactic ISM using the archive data from the CGPS at $1.42~ \rm GHz$.\footnote{The CGPS data is obtained from the Canadian Astronomy Data Centre: https://www.cadc-ccda.hia-iha.nrc-cnrc.gc.ca/en/cgps/.} The CGPS is a project involving radio, millimeter, and infrared surveys of the Galactic plane to provide arcminute-scale images of all major components of the ISM over a large part of the Galactic disk \citep{Taylor2003}. The synchrotron radio surveys are carried out at the Dominion Radio Astrophysical Observatory (DRAO). The DRAO Synthesis Telescope surveys have imaged a $73^{\degr}$ section of the Galactic plane between April 1995  and June 2000. The surveys cover the region with the longitude range of $74\fdg2<l<147\fdg3$ and the latitude extent of $-3\fdg6<b<+5\fdg6$. The full area of the CGPS is covered by 36 mosaics, each mosaic of which has a resolution of $1024\times1024$ pixels corresponding to the $5\fdg12 \times 5\fdg12$ region on the POS. 

We explore the ISM turbulence intermittency by extracting the resolution of $512\times512$ pixels from the $1024\times1024$ mosaic image to avoid the margin of images. As a representative example, we firstly provide PDFs from four mosaic images with the coordinate information listed in Table \ref{table:2}, as shown in Fig. \ref{fig:real_observation_PDF}. In practice, we filter noise-like structures of data by a Gaussian kernel of $\sigma=2$ pixels. From this figure, we can see that PDFs at different angle scales present different deviation levels of two tails from the Gaussian distribution. As the angle scale decreases, the deviation from the normal distribution increases, revealing an increment of intermittency. Moreover, it is found that the PDFs in four scenarios exhibit the presence of intermittency; which scenario has more abundant intermittency still needs to further be explored.

Next, we will quantitatively explore the degree of intermittency in four scenarios.
Fig. \ref{fig:synchrotron_k_s_four_regime_CGPS} 
 (a) depicts the kurtosis of SPI as a function of the separation angle for four sets of CGPS data. It can be seen that the kurtosis of SPI for MA2 and MC2 varies notably with the separation angle, while the kurtosis for MB2 and ME2 changes slowly. In this regard, we conclude that the SPI for MA2 and MC2 displays more intermittency than that for MB2 and ME2.
Fig. \ref{fig:synchrotron_k_s_four_regime_CGPS} 
 (b) presents the relation between the scaling exponent and order for the same data sets. It clearly shows that all the curves of the scaling exponent are nonlinear, characterising the presence of strong intermittency.
For MB2 and ME2, the profiles of the scaling exponent coincide with those of MB models, while for MA2 and MC2, the scaling exponent deviates greatly from this model. This proves that the latter is more intermittent than the former. Compared with the results of synthetic observations (see Section \ref{sec:numerical_results}), we predict that the ISM around the Galactic plane corresponds to the sub-Alfv\'enic and supersonic regimes (\citealt{Falceta2008, Heyer2008, Burkhart2009}).

\section{Discussion} \label{sec:discussion}

In this paper, we mainly explore the intermittency of MHD turbulence by the SPI and RM statistics. At sufficiently high frequencies, SPI statistics can capture the intermittency of the projected magnetic fields $B_{\perp}$ on the plane of the sky, while at low frequencies, it can provide insights into the intermittency properties of more physical quantities such as $B_{\parallel}$, $B_{\perp}$, and $n_{\rm e}$. At lower frequencies, the intermittency of SPI can also be affected by the noise-like structures. On the other hand, RM statistics can directly reflect the total intermittency for both $n_{\rm e}$ and $B_{\parallel}$, without involving the effect of $B_{\perp}$. We also tested the contribution of $n_{\rm e}$ and $B_{\parallel}$ to the intermittency of RM separately, and found that the former contributes more than the latter.

We have utilized three common statistical methods --- the PDFs, the kurtosis, and the scaling exponent of the multi-order structure function --- to explore the intermittency of MHD turbulence. The PDFs act as an indicator for the presence of intermittency when they deviate from a Gaussian distribution. Note that this method can only qualitatively reflect the intermittency level at a certain separation scale. Differently, the kurtosis and scaling exponent can provide a quantitative estimation of intermittency. The former can display the intermittency over all the separation scales. If the kurtosis varies faster with the separation scale, it implies that the fluid-structure is more intermittent. 
For the latter, when the relation between the scaling exponent and the order $p$ becomes nonlinear, this means the multifractal feature of the fluctuations and the presence of intermittency. Our studies demonstrated that the results from the three methods explored are self-consistent, the synergy of which can provide a more comprehensive understanding of MHD turbulence intermittency.  

This work is carried out in the framework of the modern understanding of MHD turbulence theory \citep{Goldreich1995}. Considering that the magnetic field and velocity retain the same cascade properties, we use three theoretical models related to velocity, i.e., the Kolmogorov, SL, and MB models, to characterise the intermittency levels of magnetic turbulence. We also tested the intermittency of the 3D magnetic field and velocity and found that they have slightly stronger intermittency than that revealed by the statistics of the SPI. Therefore, we speculate that the measured intermittency should be slightly weaker than the underlying MHD turbulence intermittency. We think that the projection effect, i.e., integration along the LOS, attenuates the intrinsic intermittency amplitude of the Galactic ISM. Similarly, the direct numerical simulations also claimed that the 3D simulation shows more intermittency than the 2D one (e.g., \citealt{Schmidt2008, Brunt2003, Brunt2004}); the latter is a projection from the 3D case.

Compared to the results from sub-Alfv{\'e}nic and subsonic turbulence (see upper panels of Fig. \ref{fig:synchrotron_three_mode}), our results demonstrated that in the case of sub-Alfv{\'e}nic and supersonic (with the presence of shocks) turbulence (see lower panels of Fig. \ref{fig:synchrotron_three_mode}), the intermittency of solenoidal mode is intensified. At the same time, we also see the intensification from intermittency of compressive (slow and fast) modes.  Note that these results we obtained are limited in the framework of solenoidal driving used in this paper, which would cause more kinetic energy to be in solenoidal motions (e.g., \citealt{Federrath2011}). In any case, this suggests that shocks may be a source of intermittency.

Earlier studies claimed that the change of spectral index only affects the amplitude of the structure function, which is only limited to the second-order structure function (\citealt{Lazarian2016, Zhang2018}). In this paper, we find that the change of spectral index ($\alpha$) affects not only the amplitude but also the scaling exponent for higher-order structures. As the spectral index increases, the intermittency of SPI becomes abundant. This is a new point. For the studies of other cases, we set the spectral index $\alpha=-1$ for the calculation of SPI. Theoretically, one expects a bottleneck effect in the power spectra of the velocity and magnetic field at large wavenumbers. However, the power spectra obtained by our data cubes do not significantly show this effect (\citealt{Wang2022, Kowal2010}). This effect does not influence the measurement of intermittency. 
However, the numerical dissipation at the
smallest resolved spatial length $0.2~\rm pc$ affects the results.

For our numerical studies, we provided the results up to the order $p=8$. We found that with increasing the order, the distributions of the scaling exponent are self-similar extending, that is, the increase in the order does not change our numerical results. However, when we use the higher order ($p>12$), there are significant fluctuations due to the limitation of the numerical resolution. For the realistic observational data, the maximum order of the structure function is only taken to the 6th order. When the 8th order is reached, the results will show abnormal fluctuations due to the denoising of the real data.

Recently, the properties of MHD turbulence have been studied using the synchrotron polarization statistics (see \citealt{Zhang2022} for a recent review), including the spatial and frequency analysis techniques \citep{Lazarian2016,Zhang2016,Lee2016}, gradient techniques (e.g., \citealt{Lazarian2018,Zhang2019}), and quadrupole ratio modulus (\citealt{Lee2019,Wang2020}). It is stressed that these works mainly focused on the inertial range of turbulence cascade. Differently, our current work covers a wide range of spatial scales, particularly involving the small-scale non-noise structure, to understand the properties of compressible MHD turbulence.

\cite{Cho2010} proposed that analysing intermittent features can separate foreground signals from cosmic microwave background signals via the high-order structure function. The intermittency can also explain the observed strong and rapid variations from pulsar magnetosphere \citep{Zelenyi2015}. We expect that measuring intermittency may distinguish the difference between \cite{Goldreich1995} and \cite{Boldyrev2006} theories, which will be discussed elsewhere.

\section{Summary} \label{sec:conclusion}
Using real observational data from the CGPS together with MHD turbulence simulation, we have investigated how to recover the intermittency of the magnetized ISM. The main results are briefly summarised as follows.
\begin{itemize}
\item The SPI statistics can be used to probe the intermittency of MHD turbulence. The most significant intermittency appears in the sub-Alfv{\'e}nic and supersonic turbulence regime, while the least intermittency in the sub-Alfv{\'e}nic and subsonic one.

\item The intermittency measured by the SPI depends on the level of the Faraday depolarization. The intermittency measured by the RM shows a strong dependence on the sonic Mach number, with significant intermittency occurring in the supersonic turbulence regime. Therefore, RM statistics can recover the intermittency of thermal electron density and magnetic field component along the LOS.

\item Slow mode dominates the intermittency of MHD turbulence in the sub-Alfv\'enic turbulence regime, where Alfv\'en (for supersonic) and fast (for subsonic) modes almost present a negligible intermittency. 

\item With realistic observations from the CGPS, we find that the Galactic ISM at the low latitude region corresponds to the sub-Alfv\'enic and supersonic turbulence regime.

\end{itemize}

\begin{acknowledgements}
We would like to thank the anonymous referee for constructive comments that have significantly improved our manuscript. J.F.Z. is grateful for the support from the National Natural Science Foundation of China (Nos. 12473046 and 11973035), the Hunan Natural Science Foundation for Distinguished Young Scholars (No. 2023JJ10039), and the China Scholarship Council for the overseas research fund. F.Y.X. thanks the support from the National Natural Science Foundation of China (grant No. 12373024). R.Y.W. is grateful for the support from the Xiangtan University Innovation
Foundation For Postgraduate (No. XDCX2022Y071).
\end{acknowledgements}

\bibliography{ms}  
\bibliographystyle{aa}

\begin{appendix}
\section{Multi-order structure functions}
\begin{figure*}
\centering
\includegraphics[width=1.0\textwidth]{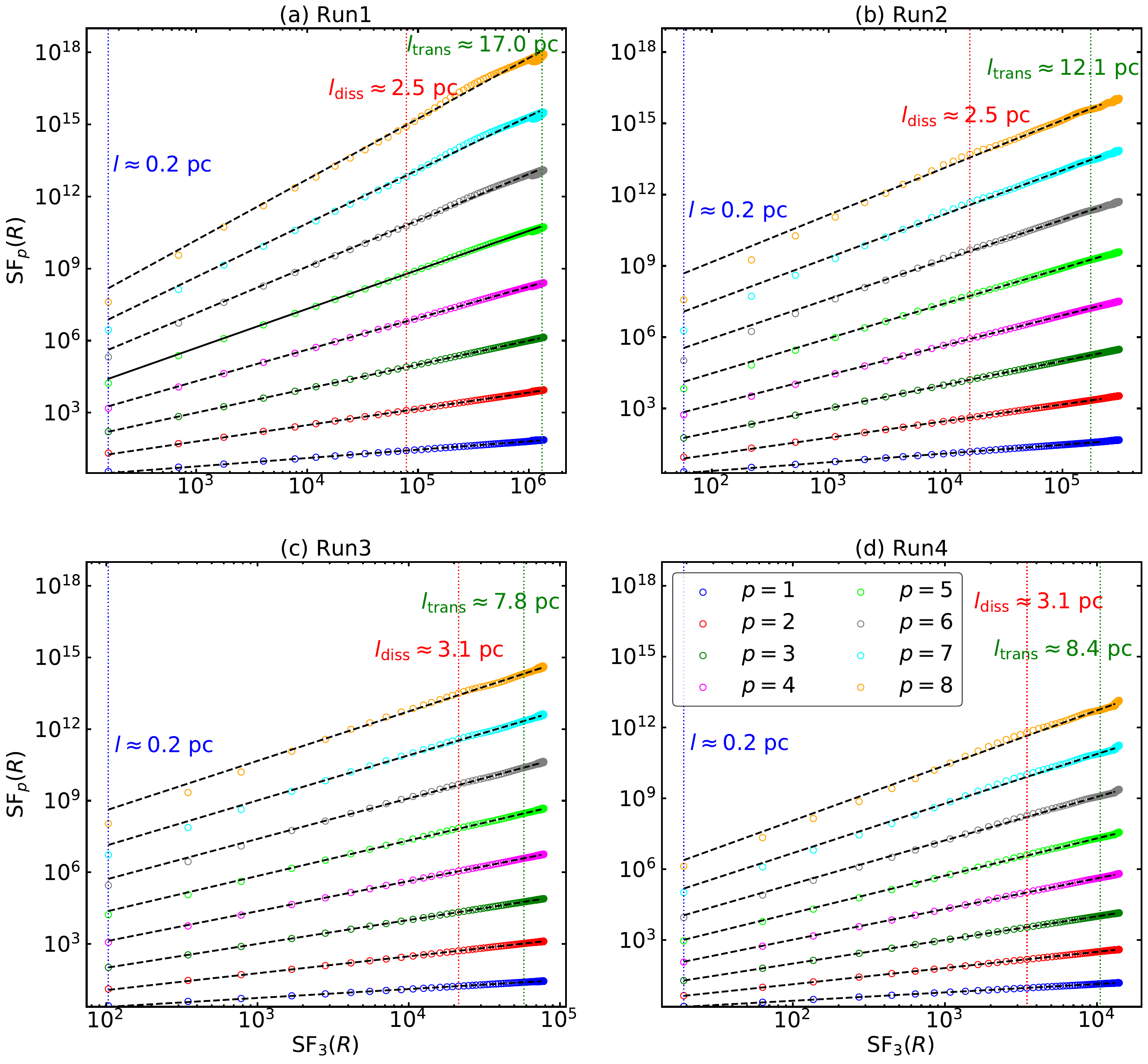}
\caption{Structure functions of the SPI with different orders (from $p=1$ to 8) as a function of ${\rm SF}_{3}(R)$ under the extended self-similarity hypothesis. The vertical blue, green, and red dotted lines denote the values of third-order structure function in the smallest resolved spatial length, transition scale, and dissipation scale, respectively. The black dashed lines show a linear fit to the structure function on log-log scales.
}\label{fig:appendix_SPI}
\end{figure*}

\begin{figure*}
\centering
\includegraphics[width=1.0\textwidth]{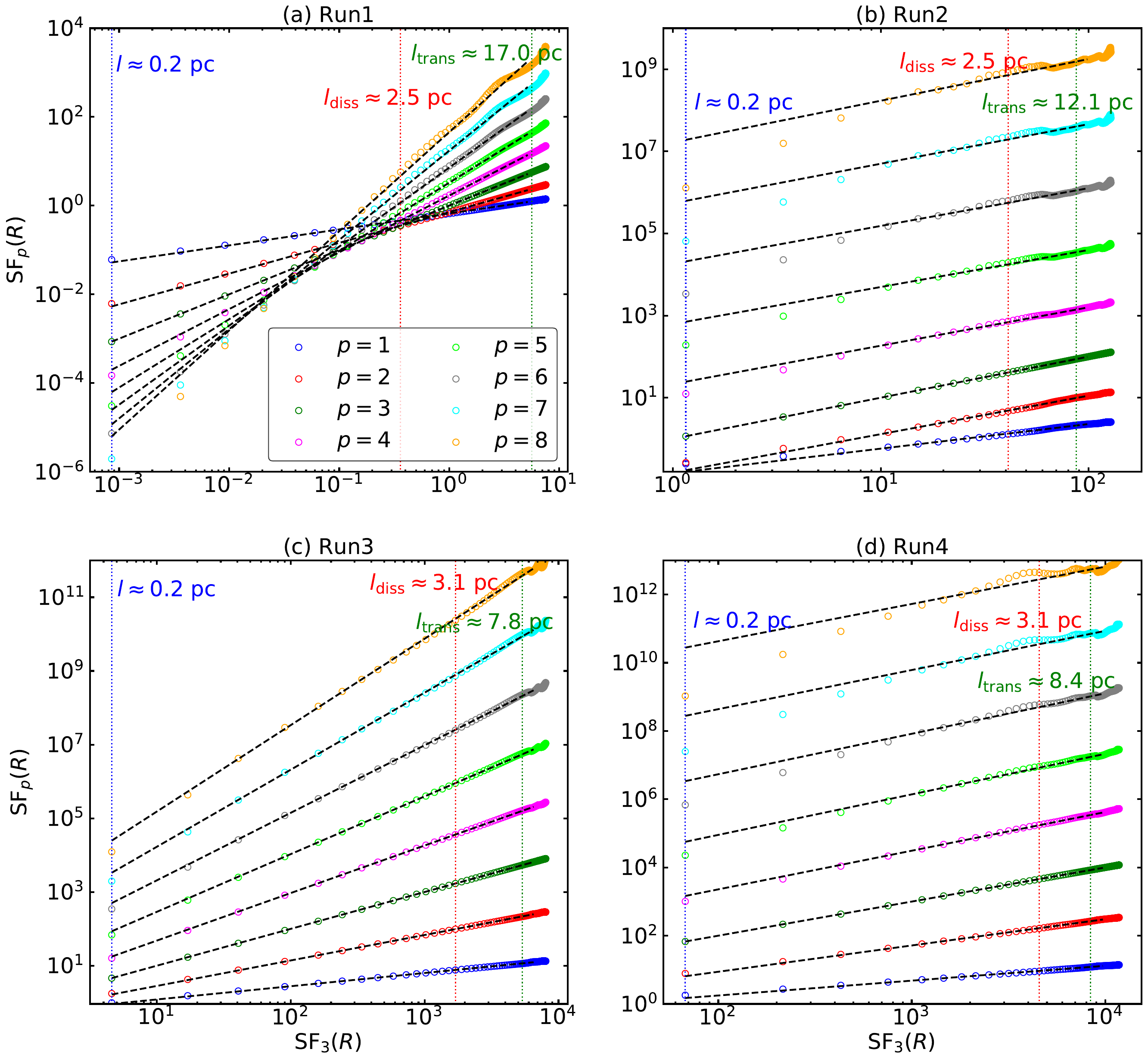}
\caption{Same as Fig. \ref{fig:appendix_SPI}, but for the multi-order structure functions of the RM statistics.
}\label{fig:appendix_RM}
\end{figure*}

In this appendix, we provide the multi-order structure functions of the SPI and RM in Figs. \ref{fig:appendix_SPI} and \ref{fig:appendix_RM}, respectively. These figures exhibit the structure functions with different orders as a function of ${\rm SF}_{3}(R)$. As is shown in these figures, we can see a good power-law relation. For quantitative characterization, we perform a linear fitting process in the interval between $R=0.6$ and $R=15.6~\rm pc$. According to the fitting from these two figures, we can further obtain the results of Figs. \ref{fig:synchrotron_scaling} (c) and \ref{fig:faraday_k_s_four_regime}, respectively, where the uncertainties arising from the linear fitting are reflected by the error bars, i.e., the standard deviations.

\end{appendix}
\end{document}